\documentclass{article}

\usepackage{arxiv}

\usepackage[utf8]{inputenc} 
\usepackage[T1]{fontenc}    
\usepackage{hyperref}       
\usepackage{url}            
\usepackage{booktabs}       
\usepackage{amsfonts}       
\usepackage{nicefrac}       
\usepackage{microtype}      
\usepackage{lipsum}
\usepackage{graphicx}
\usepackage{wrapfig}
\usepackage{amsthm}
\usepackage[ruled,linesnumbered]{algorithm2e}
\usepackage[numbers]{natbib}

\newtheorem{example}{Example}

\title{Probabilistic Counting in Uncertain Spatial Databases using Generating Functions}

\author{
 Andreas Z\"ufle \\
  George Mason University\\
  \texttt{azufle@gmu.edu} \\
  }

\begin{document}
\maketitle

\begin{abstract}
Location data is inherently uncertain for many reasons including 1) imprecise location measurements, 2) obsolete observations that are often interpolated, and 3) deliberate obfuscation to preserve location privacy. 
What makes handling uncertainty data challenging is the exponentially large number of possible worlds, which lies in $O(2^N)$, for a database having $N$ uncertain objects as it has been shown that general query processing in uncertain spatial data is NP-hard.
Many applications using spatial data require counting the number of spatial objects within a region. An example is the $k$-Nearest Neighbor ($k$NN) query: Asking if an object $A$ is a $k$NN of another object $Q$ is equivalent to asking whether no more than $k-1$ objects are located inside the circle centered at $Q$ having a radius equal to the distance between $Q$ and $A$. 

For this problem of counting uncertain objects within a region, an efficient solution based on Generating Functions has been proposed and successfully used in many applications, including range-count queries, $k$NN queries, distance ranking queries, and reverse $k$NN queries. This spatial gem describes the generating function technique for probabilistic counting and provides examples and implementation details.
\end{abstract}


\section{Introduction}
Our ability to unearth valuable knowledge from large sets of spatial data is often impaired by the uncertainty of the data which geography has been named the ``the Achilles heel of GIS'' \cite{goodchild1998uncertainty} for many reasons: 1) Imprecision is caused by physical limitations of sensing devices and connection errors, 2) Data records may be obsolete; 3) Data can be obtained from unreliable sources, such as volunteered geographic information; and 4) Data may be deliberately obfuscated to preserve the privacy of users.  These issues introduce the notion of uncertainty in the context of spatio-temporal data management. Many algorithms have been proposed in the last decade to handle different spatial query predicates (such as distance range, $k$NN, and distance ranking) described in various tutorials~\cite{renz2010similarity,cheng2014managing,zufle2017handling,zufle2020managing} and surveys~\cite{zufle2021uncertain,aggarwal2008survey}. A commonly used technique that allows many of these algorithms to run efficiently leverages the technique of Generating Functions to efficiently aggregate an exponential number of possible worlds in polynomial time. This technique is described, along with examples and implementation, in this spatial gem.

\begin{figure}[t]
    \centering
    \includegraphics[width=0.7\columnwidth]{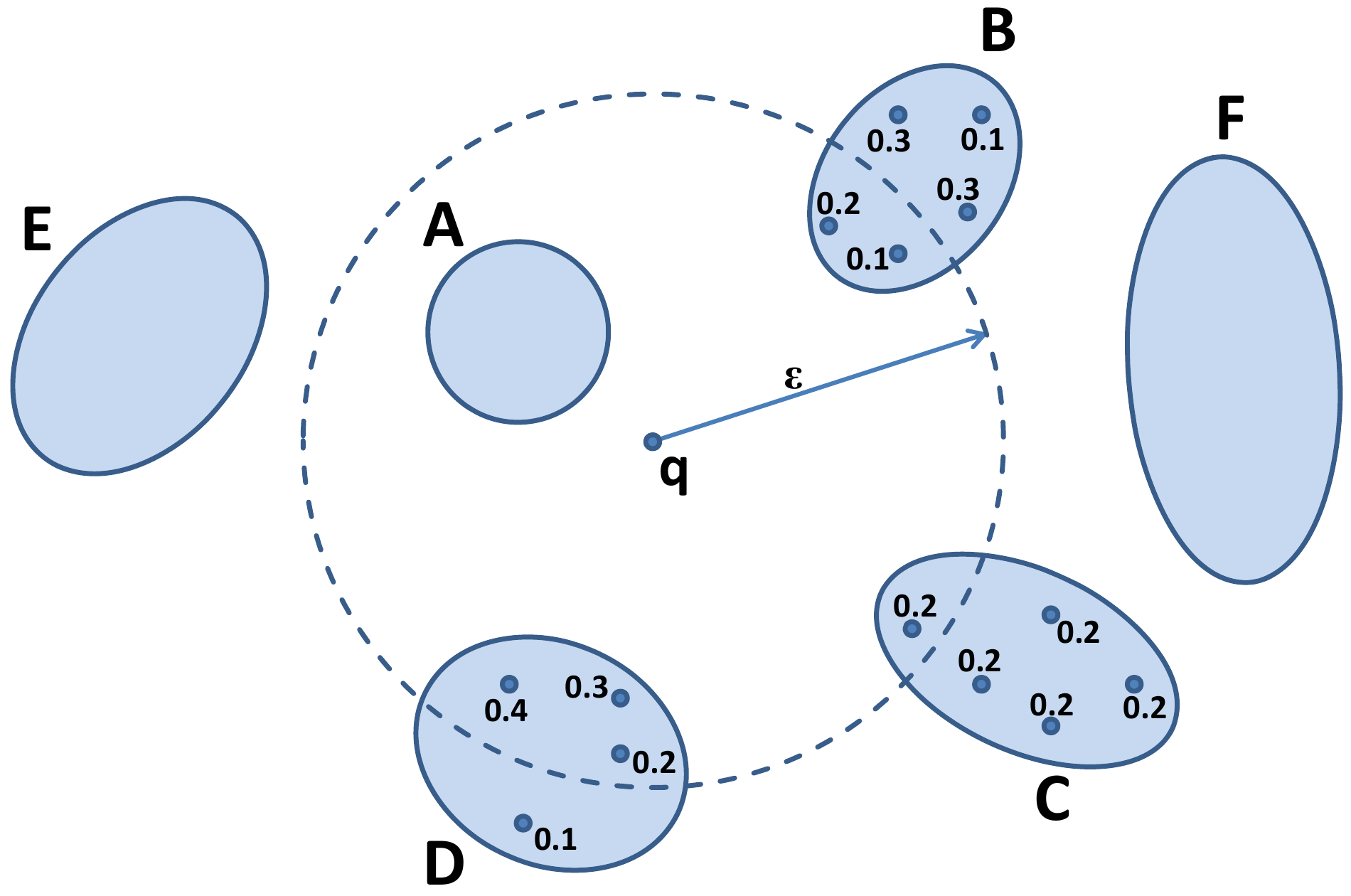}
    \caption{Example of an uncertain $\epsilon$-range query. Object $A$ is a true hit, objects $B$, $C$ and $D$ are possible hits.}
    \label{fig:blume}
\end{figure}

An example of an uncertain (toy) database is shown in Figure~\ref{fig:blume} having six uncertain objects $\{A,B,C,D,E,F\}$. Rather than having a single unique (crisp) location, an object in an uncertain database may have multiple alternatives, each associated with a corresponding probability of being the true location. For example, object $B$ has five possible locations and object $D$ has four possible locations. There exist multiple models for uncertain data, either describing uncertain objects by discrete (and finite) sets of alternatives, or by describing uncertain objects by continuous distributions of (uncountably infinite) possible locations~\cite{zufle2021uncertain}. The most prominent systems for uncertain relational data management are MayBMS
\cite{antova2008fast}, MystiQ \cite{boulos2005mystiq}, Trio \cite{agrawal2006trio}, and BayesStore \cite{wang2008bayesstore} which allow to efficiently answer traditional queries that select subsets of data based on predicates or join different datasets based on conditions. While these existing systems efficiently support simple projection-selection-join queries, they offer no support for complex queries and data mining tasks. A likely reason for this gap
is the theoretic result of \cite{dalvi2007efficient} which shows that the general problem of query processing in uncertain databases is \#P-hard in the number of database objects.
%
%
%

While this result implies that general query processing on uncertain data is hard, it does not outrule the possibility of efficient solutions for specific query types. And in fact, many important classes of spatial queries have efficient (polynomial time) solutions, including range count queries~\cite{follmann2011continuous}, nearest neighbor queries \cite{ReyKalPra04,IjiIsh09,ChengCMC08}, k-nearest neighbor queries \cite{BesSolIly08,LjoSin07,CheCheCheXie09} and (similarity-) ranking queries \cite{CorLiYi09,SolIly09,li2009unified,cormode2009semantics,li2010ranking,li2009unified,hua2008ranking,yi2008efficient}.

All these classes of spatial queries have in common that they count the number of spatial objects that fall within a region. A range count query directly returns the distribution of the number of objects within a specified query region. To decide if an object $A$ is a $kNN$ of a query object $Q$; a $kNN$ query computes that probability that less than $k$ objects are closer to $Q$ than $A$, thus counting a number of objects within a distance of less than the distance between $Q$ and $A$; and for a distance ranking query, the probability that an object $A$ has the $k$-th nearest objects of $Q$ is the probability that exactly $k-1$ objects (other than $A$) have a distance to $Q$ less than the distance between $Q$ and $A$.

\begin{example}
As an example of counting the number of uncertain objects within a region, reconsider Figure~\ref{fig:blume}, and assume a query that counts the number of uncertain objects within a distance of $\epsilon$ from a query object $q$. We first note that this query answer is a random variable, which depends on the locations of the uncertain objects (which are random variables, too). We observe that objects $E$ and $F$ are guaranteed to be outside the range, such that we can prune them from our computation. We also note that object $A$ is guaranteed to be in the range, allowing us to increment the query result by one without having to further consider this object. For objects $B$, $C$, and $D$, the events of being located inside the query range are random variables with probabilities of $0.3$, $0.2$, and $0.9$, respectively. Thus, the result of this query is a random variable having a sample space of $\{1,2,3,4\}$, and mapping each of these possible results to their probability. For example, the probability of having exactly one object in the range is $0.7\cdot 0.8 \cdot 0.1=0.056$. For the probability that exactly two objects are inside the range, we can add the probabilities on the three possible worlds where exactly one object out of $\{B,C,D\}$ is inside the range. 
\end{example}
In the general case of computing the probability that exactly $k$ out of $n$ uncertain objects are inside the query range, we need to aggregate the probabilities of ${n \choose k}$ combinations of objects to be inside the query range. Straightforward approaches which enumerate all the ${n \choose k}$ possible worlds the number of which is in $O(n^k)$. 

Yet, for this problem of counting the (distribution of the) number of uncertain objects within a query range two efficient solutions based on 1) the Poisson-Binomial Recurrence~\cite{yi2008efficient,hua2008ranking,bernecker2010scalable} and 2) based on Generating Functions~\cite{li2009unified} have been proposed independently in the literature. These solutions allow to aggregate the probabilities of an exponential number of possible combinations of objects in polynomial time, allowing us to answer many important spatial query types efficiently. This spatial gem describes how the generating function technique, which was first presented in the context of distance ranking by Li, Saha, and Deshpande in the best paper of VLDB 2009, can be used to efficiently answer spatial queries on uncertain data.

\section{Generating Functions for Probabilistic Counting}
Let $\mathcal{X}=\{X_1,...,X_N\}$ denote the set of objects having a non-zero probability of being located in the query region and let $p_i$ denote the probability of object $X_i$ to be located inside the query region. We can model each $X_i\models B(p_i)$ as a Bernoulli distributed random variable that has a probability of $p_i$ of being $1$ and a probability of $(1-p_i)$ of being $0$. With this model, the count of objects inside the query region is the sum $\sum_{i=1}^N B(p_i)$. We note that since the probabilities $p_i$ are not identical this random variable does not follow a Binomial distribution, but is instead known as a Poisson-Bionomial distribution having parameters $\{p_1,...,p_N\}$~\cite{hua2008ranking}.
Our goal is to evaluate this random variable $\sum_{i=1}^N B(p_i)$ efficiently, that is, for each $0 \leq k \leq N$ we want to derive the probability $P(\sum_{i=1}^N B(p_i)=k)$.

For this purpose, represent each random variable $X_i$ by a polynomial
$poly(X_i)=p_i\cdot x+(1-p_i)$. Consider the generating function
\begin{equation}\label{eq:genfkts}
\mathcal{F}^N=\prod_{i=1}^N poly(X_i)=\sum_{i=0}^{N} c_ix^i.
\end{equation}
The coefficient $c_i$ of $x^i$ in the expansion of $\mathcal{F}^N$
equals the probability $P(\sum_{n=1}^N X_n=i)$ (\cite{li2009unified}). For
example, the monomial $0.25\cdot x^4$ implies that with a
probability of $0.25$, the sum of all Bernoulli random variables
equals four.

The expansion of $N$ polynomials, each containing two monomials
leads to a total of $2^N$ monomials, one monomial for each
sequence of successful and unsuccessful Bernoulli trials, i.e.,
one monomial for each possible world. To reduce this complexity,
an iterative computation of $\mathcal{F}^N$, can be used, by
exploiting that
\begin{equation}\label{eq:genfkts2}
\mathcal{F}^k=\mathcal{F}^{k-1}\cdot poly(X_k).
\end{equation}
This rewriting of Equation \ref{eq:genfkts} allows to inductively
compute $\mathcal{F}^k$ from $\mathcal{F}^{k-1}$. The induction is
started by computing the polynomial $\mathcal{F}^0$, which is the
empty product which equals $1$, the neutral element of multiplication,
i.e., $\mathcal{F}^0=1$. To understand the semantics of this
polynomial, the polynomial $\mathcal{F}^0=1$ can be rewritten as
$\mathcal{F}^0=1\cdot x^0$, which we can interpret as the
following tautology:``with a probability of one, the sum of all
zero Bernoulli trials equals zero.'' After each iteration, we can
unify monomials having the same exponent, leading to a total of at
most $k+1$ monomials after each iteration. This unification step
allows to remove the combinatorial aspect of the problem, since
any monomial $x^i$ corresponds to a class of equivalent worlds,
such that this class contains only and all of the worlds where the
sum $\sum_{k=1}^N X_k=1$. In each iteration, the number of these
classes is at most $k$ and the probability of each class is given by the
coefficient of $x^i$.

\begin{example}
As an example, consider again the running example of Figure~\ref{fig:blume}. For each object, we first obtain the probability of being located inside the query region (which can be done in linear time using a range query and aggregating the probabilities of instances inside the query region). For the four objects $A$, $B$, $C$, $D$, $E$, and $F$, we obtain probabilities of being inside the query region of $1.0$, $p_1:=0.3$, $p_2:=0.2$, $p_3:=0.9$, $0$, and $0$, respectively. We can safely prune objects $E$ and $F$ since they cannot affect the query result. We can also prune object $A$ by increasing the result by $1$ since we know $A$ must be inside the query range. Given the probabilities $p_1=0.3$, $p_2=0.2$, and $p_3=0.9$ we obtain the three generating polynomials
$poly(X_1)=(0.3x+0.7)$, $poly(X_2)=(0.2x+0.8)$, and
$poly(X_3)=(0.9x+0.1)$. We trivially
obtain $\mathcal{F}^0=1$. Using Equation \ref{eq:genfkts2} we get\vspace{-0.0cm}
$$
\mathcal{F}^1=\mathcal{F}^{0}\cdot poly(X_1)=1\cdot (0.3x +
0.7)=0.3x + 0.7.
$$\vspace{-0.0cm}
Semantically, this polynomial implies that out of the first one
Bernoulli trials, the probability of having a sum of one is
$0.3$ (according to monomial $0.3x=0.3x^1$), and the probability of
having a sum of zero is $0.7$ (according to monomial $0.7=0.7x^0$.
Next, we compute $F^2$, again using Equation \ref{eq:genfkts2}:\vspace{-0.0cm}
$$
\mathcal{F}^2=\mathcal{F}^{1}\cdot poly(X_2)=(0.3x^1 +
0.7^0)\cdot
(0.2x^1+0.8x^0)
=
$$
$$
0.06x^1x^1+0.24x^1x^0+0.14x^0x^1+0.56x^0x^0
$$\vspace{-0.0cm}
In this expansion, the monomials have deliberately not been
unified to give an intuition of how the generating function
technique is able to identify and unify equivalent worlds. In the
above expansion, there is one monomial for each possible world.
For example, the monomial $0.14x^0x^1$ represents the world where
the first trial was unsuccessful (represented by the $0$ in the
first exponent) and the second trial was successful (represented by
the $1$ in the second exponent). The above notation allows to
identify the sequence of successful and unsuccessful Bernoulli
trials, clearly leading to a total of $2^k$ possible worlds for
$\mathcal{F}^k$. However, we know that we only need to compute the
total number of successful trials, we do not need to know the
sequence of successful trials. Thus, we may treat worlds
having the same number of successful Bernoulli trials
equivalently, to avoid the enumeration of an exponential number of
sequences. This is done implicitly by polynomial multiplication,
exploiting that\vspace{-0.0cm}
$$
0.06x^1x^1+0.24x^1x^0+0.14x^0x^1+0.56x^0x^0=0.06x^2+0.24x^1+0.14x^1+0.56x^0
$$\vspace{-0.0cm}
This representation no longer allows to distinguish the sequence
of successful Bernoulli trials. This loss of information is
beneficial, as it allows to unify possible worlds having the same
sum of Bernoulli trials
$$
0.06x^2+0.24x^1+0.14x^1+0.56x^0=0.06x^2+0.38x^1+0.56x^0
$$
The remaining monomials represent an equivalence class of possible
worlds. For example, monomial $0.38x^1$ represents all worlds
having a total of one successful Bernoulli trial out of the first two trials. This is
evident since the coefficient of this monomial was derived from
the sum of both worlds having a total of one successful Bernoulli
trial. In the next iteration, we compute:
$$
\mathcal{F}^3=\mathcal{F}^{2}\cdot
poly(X_3)=(0.06x^2+0.38x^1+0.56x^0)\cdot (0.9x+0.1)
$$
$$
=0.054x^2x^1+0.006x^2x^0+0.342x^1x^1+0.038x^1x^0+0.504x^0x^1+0.056x^0x^0
$$
This polynomial represents the three classes of possible worlds in
$\mathcal{F}^2$ combined with the two possible results of the
third Bernoulli trial, yielding a total of $3\dot2$ monomials.
Unification yields
$$
0.054x^2x^1+0.006x^2x^0+0.342x^1x^1+0.038x^1x^0+0.504x^0x^1+0.056x^0x^0=
$$
$$
0.054x^3+0.348x^2+0.542x^1+0.056x^0
$$
This polynomial describes the PDF of $\sum_{i=1}^3 X_i$ (having $X_1=B, X_2=C, X_3=D$), since
each monomial $c_ix^i$ implies that the probability, that out of
all three Bernoulli trials, the total number of successful events
equals $i$, is $c_i$. Thus, we get $P(\sum_{i=1}^3 X_i=0)=0.0056$,
$P(\sum_{i=1}^3 X_i=1)=0.542$, $P(\sum_{i=1}^3 X_i=2)=0.348$, and
$P(\sum_{i=1}^3 X_i=3)=0.054$.
\end{example}

\section{Complexity Analysis}
The generating function technique requires a total of $N$
iterations (as in the worst case, all uncertain objects have a non-zero non-one probability of being in the query region). 
In each iteration $1\leq k \leq N$, a polynomial of
degree $k-1$, and thus of maximum length $k$, is multiplied with a
polynomial of degree $1$, thus having a length of $2$. This
requires to compute a total of $(k+1)\cdot 2$ monomials in each
iteration, each requiring a scalar multiplication. This leads to a
total time complexity of $\sum_{i=1}^N 2k+2\in O(N^2)$ for the
polynomial expansions. Unification of a polynomial of length $k$
can be done in $O(k)$ time, exploiting that the polynomials are
sorted by the exponent after expansion. Unification at each
iteration leads to a $O(n^2)$ complexity for the unification step.
This results in a total complexity of $O(n^2)$, similar to the
Poisson binomial recurrence approach.

Many spatial query predicates do not require to compute the full probability mass of the distribution of the number of objects within the query range but only require the probability of having less or equal than a specified parameter $K$ of objects in the query range. For example, to find the probability that an object is among the $K$-nearest neighbors of a query object, it is sufficient to compute the probability that at most $K-1$ objects are closer (within a shorter range). In this case, all monomials having an exponent greater or equal to $K$ can be pruned from the computation. In this case, where the length of the expanded polynomial in each iteration is bounded by $K$, thus yielding a run-time complexity of $O(k\cdot n)$. This efficient computation can also be leveraged for the case of distance ranking, where the challenge is to find the probability that exactly $K-1$ other objects are closer to a query object for an object to be exactly the $K$'th nearest neighbor (i.e., having a distance rank of $K$). This task requires only to find the coefficient $c_{K-1}$ of the expanded monomial $c_{K-1}x^{K-1}$ having an exponent of $K-1$. Since in each iteration of multiplying and expanding monomials the coefficient $c_{K-1}$ only depends on the coefficients $c_{K-2}$ and $c_{K-1}$ of previous iterations, we may also discard monomials having an exponent of $K$ or greater to answer distance ranking queries.

To summarize, using generating functions we can compute the distribution of the number of objects within a query range in $O(n^2)$, where $n$ is the number of database objects. In cases such as $K$NN or distance ranking queries where we only need to know the probability of having at most or exactly $K$ objects within the query range, we can reduce this complexitiy to $O(K\cdot n)$ by truncating intermediate polynomials.

\section{Implementation}
A Python implementation can be found in the following GitHub repository \url{https://github.com/azufle/generating_functions}. This implementation, both as a Jupyter notebook and a classic .py script defines a function that efficiently computes the probability mass function of a probabilistic count given a list of probabilities. Additional documentation can be found in the repository.

\section{Variants, Extensions, and Improvements}
This section surveys a variety of extensions and improvements of the classic generating functions.

\subsection{Acceleration using Discrete Fourier Transform}
An advantage of the generating function approach is that this
naive polynomial multiplication can be accelerated using Discrete
Fourier Transform (DFT). This technique allows to reduce to total
complexity of computing the sum of $N$ Bernoulli random variables
to $O(N log^2 N)$ (\cite{LiSahDes11}). This acceleration is
achieved by exploiting that DFT allows to expand two polynomials
of size $k$ in $O(k log k)$ time. Equi-sized polynomials are
obtained in the approach of \cite{LiSahDes11}, by using a divide
and conquer approach, that iteratively divides the set of $N$
Bernoulli trials into two equi-sized sets. Their recursive
algorithm then combines these results by performing a polynomial
multiplication of the generating polynomials of each set. More
details of this algorithm can be found in \cite{LiSahDes11}.

\subsection{Extension to Uncertain Counts}
In many applications, a probabilistic event may not only have two possible outcomes (Yes/No, Success/Failure), but may have a third outcome that represents an unknown/undecided/uncertain state. For example, given incomplete trajectories of objects and their resulting uncertainty regions at a time (for example, described by a bounding box of possible locations), there may be possible worlds where 1) an object is within the query region, 2) an object is outside the query region, and 3) containment of the object within the query region cannot be decided due to uncertainty. 

For such cases, the addition of an unknown state has been proposed to be represented by generating function in~\cite{bernecker2011novel}. For each object $X_i\in\mathcal{X}=\{X_1,...,X_N\}$ having a probability of $p_i$ to satisfy the query condition (such as being located within the query region), a probability of $\bar{p}_i$ to not satisfy the query condition, and a probability of $1-p_i-\bar{p}_i$ of being in an unknown state, we can consider the generating function:
$$
\mathcal{F}^N=\prod_{i=1}^N p_i\cdot x + (1-p_i-\bar{p}_i)\cdot y + \bar{p}_i
$$
Intuitively, the anonymous variable $x$ denotes an event of satisfying the query condition and the anonymous variable $y$ denotes the event of an undecided satisfaction of the query condition. In the expanded polynomial, a monomial such as $c_{i,j} x^iy^j$ corresponds to a possible world having a probability of $c_{i,j}$ of having $i$ objects guaranteed to satisfy the query condition and $j$ additional objects possibly satisfying the query condition. For example, a monomial $0.13x^3y^2$ corresponds to a possible world having a probability of $0.13$ and having at least $3$ but no more than $3+2=5$ objects satisfy the query predicate.

\subsection{Dynamic Polynomials}
In many applications, the probability of an object to be inside a query range may change dynamically. For example, mobile objects may send update location information to change their uncertainty region. In this case, the probability distribution of the number of objects inside a query range changes as well. To update the probability distribution, we may recompute from scratch, using all objects having non-zero probability of being inside the query range. However, such an approach may be inefficient when there is a large number of such objects having frequent updates. To update the probability distribution of a probabilistic count, we can use polynomial division as described in~\cite{hubig2012continuous}. Thus, having a database $\mathcal{X}=\{X_1,...,X_N\}$ of objects each having a probability of $p_i$ to be inside the query and given the polynomial $\mathcal{F}^N$ that describes the probability distribution of the number of objects inside the query range, assume that an object $X_j$ changes it's probability from $p_j$ to $p_j^\prime$. We can update the $\mathcal{F}^N$ by removing the old effect of the old probability $p_i$ through polynomial division of polynomial $poly(X_j)= p_i\cdot x + 1 - p_i$ and by including the effect of the effect of the new probability through multiplication with polynomial $poly(X_j^\prime)= p_i^\prime \cdot x + 1 - p_i^\prime$. Combining both steps, we obtain the updated polynomial $\mathcal{F}^{\prime N}$:
$$
\mathcal{F}^{\prime N} =\mathcal{F}^{N} \frac{poly(X_j^\prime)}{poly(X_j)} = \mathcal{F}^{N} \frac{p_i^\prime \cdot x + 1 - p_i^\prime}{p_i\cdot x + 1 - p_i}.
$$

\clearpage

\setlength{\bibsep}{0pt plus 0.3ex}
{\small

}



\begin{thebibliography}{10}

\bibitem{goodchild1998uncertainty}
Michael~F Goodchild.
\newblock {Uncertainty: The Achilles Heel of GIS}.
\newblock {\em Geo Info Systems}, 8(11):50--52, 1998.

\bibitem{renz2010similarity}
Matthias Renz, Reynold Cheng, and Hans-Peter Kriegel.
\newblock Similarity search and mining in uncertain databases.
\newblock {\em Proceedings of the VLDB Endowment}, 3(1-2):1653--1654, 2010.

\bibitem{cheng2014managing}
Reynold Cheng, Tobias Emrich, Hans-Peter Kriegel, Nikos Mamoulis, Matthias
  Renz, Goce Trajcevski, and Andreas Z{\"u}fle.
\newblock Managing uncertainty in spatial and spatio-temporal data.
\newblock In {\em 2014 IEEE 30th International Conference on Data Engineering},
  pages 1302--1305. IEEE, 2014.

\bibitem{zufle2017handling}
Andreas Z{\"u}fle, Goce Trajcevski, Dieter Pfoser, Matthias Renz, Matthew~T
  Rice, Timothy Leslie, Paul Delamater, and Tobias Emrich.
\newblock Handling uncertainty in geo-spatial data.
\newblock In {\em 2017 IEEE 33rd International Conference on Data Engineering
  (ICDE)}, pages 1467--1470. IEEE, 2017.

\bibitem{zufle2020managing}
Andreas Z{\"u}fle, Goce Trajcevski, Dieter Pfoser, and Joon-Seok Kim.
\newblock Managing uncertainty in evolving geo-spatial data.
\newblock In {\em 2020 21st IEEE International Conference on Mobile Data
  Management (MDM)}, pages 5--8. IEEE, 2020.

\bibitem{zufle2021uncertain}
Andreas Z{\"u}fle.
\newblock Uncertain spatial data management: An overview.
\newblock {\em Handbook of Big Geospatial Data}, pages 355--397, 2021.

\bibitem{aggarwal2008survey}
Charu~C Aggarwal and S~Yu Philip.
\newblock A survey of uncertain data algorithms and applications.
\newblock {\em IEEE Transactions on knowledge and data engineering},
  21(5):609--623, 2008.

\bibitem{antova2008fast}
Lyublena Antova, Thomas Jansen, Christoph Koch, and Dan Olteanu.
\newblock {Fast and Simple Relational Processing of Uncertain Data}.
\newblock In {\em 2008 IEEE 24th International Conference on Data Engineering},
  pages 983--992. IEEE, 2008.

\bibitem{boulos2005mystiq}
Jihad Boulos, Nilesh Dalvi, Bhushan Mandhani, Shobhit Mathur, Chris Re, and Dan
  Suciu.
\newblock {MYSTIQ: A system for finding more answers by using probabilities}.
\newblock In {\em Proceedings of the 2005 ACM SIGMOD International Conference
  on Management of Data}, pages 891--893. ACM, 2005.

\bibitem{agrawal2006trio}
Parag Agrawal, Omar Benjelloun, Anish~Das Sarma, Chris Hayworth, Shubha Nabar,
  Tomoe Sugihara, and Jennifer Widom.
\newblock {Trio: A System for Data, Uncertainty, and Lineage}.
\newblock {\em Proc. of VLDB 2006 (demonstration description)}, 2006.

\bibitem{wang2008bayesstore}
Daisy~Zhe Wang, Eirinaios Michelakis, Minos Garofalakis, and Joseph~M
  Hellerstein.
\newblock {BAYESSTORE: Managing Large, Uncertain Data Repositories with
  Probabilistic Graphical Models}.
\newblock {\em Proceedings of the VLDB Endowment}, 1(1):340--351, 2008.

\bibitem{dalvi2007efficient}
Nilesh Dalvi and Dan Suciu.
\newblock Efficient query evaluation on probabilistic databases.
\newblock {\em The VLDB Journal}, 16(4):523--544, 2007.

\bibitem{follmann2011continuous}
Anna Follmann, Mario~A Nascimento, Andreas Z{\"u}fle, Matthias Renz, Peer
  Kr{\"o}ger, and Hans-Peter Kriegel.
\newblock Continuous probabilistic count queries in wireless sensor networks.
\newblock In {\em International Symposium on Spatial and Temporal Databases},
  pages 279--296. Springer, 2011.

\bibitem{ReyKalPra04}
Reynold Cheng, Dmitri~V. Kalashnikov, and Sunil Prabhakar.
\newblock Querying imprecise data in moving object environments.
\newblock {\em IEEE Trans. Knowl. Data Eng.}, 16(9):1112--1127, 2004.

\bibitem{IjiIsh09}
Yuichi Iijima and Yoshiharu Ishikawa.
\newblock Finding probabilistic nearest neighbors for query objects with
  imprecise locations.
\newblock In {\em {MDM}}, pages 52--61, 2009.

\bibitem{ChengCMC08}
Reynold Cheng, Jinchuan Chen, Mohamed~F. Mokbel, and Chi-Yin Chow.
\newblock Probabilistic verifiers: Evaluating constrained nearest-neighbor
  queries over uncertain data.
\newblock In {\em {ICDE}}, pages 973--982, 2008.

\bibitem{BesSolIly08}
George Beskales, Mohamed~A. Soliman, and Ihab~F. IIyas.
\newblock Efficient search for the top-k probable nearest neighbors in
  uncertain databases.
\newblock {\em Proc. VLDB Endow.}, 1(1):326--339, 2008.

\bibitem{LjoSin07}
Vebjorn Ljosa and Ambuj~K. Singh.
\newblock Apla: Indexing arbitrary probability distributions.
\newblock In {\em {ICDE}}, pages 946--955, 2007.

\bibitem{CheCheCheXie09}
Reynold Cheng, Lei Chen, Jinchuan Chen, and Xike Xie.
\newblock Evaluating probability threshold k-nearest-neighbor queries over
  uncertain data.
\newblock In {\em {EDBT}}, pages 672--683, 2009.

\bibitem{CorLiYi09}
G.~Cormode, F.~Li, and K.~Yi.
\newblock Semantics of ranking queries for probabilistic data and expected
  results.
\newblock In {\em 2009 IEEE 25th International Conference on Data Engineering},
  pages 305--316, 2009.

\bibitem{SolIly09}
M.A. Soliman and I.F. Ilyas.
\newblock Ranking with uncertain scores.
\newblock In {\em 2009 IEEE 25th International Conference on Data Engineering},
  pages 317--328, 2009.

\bibitem{li2009unified}
Jian Li, Barna Saha, and Amol Deshpande.
\newblock {A Unified Approach to Ranking in Probabilistic Databases}.
\newblock {\em Proceedings of the VLDB Endowment}, 2(1):502--513, 2009.

\bibitem{cormode2009semantics}
Graham Cormode, Feifei Li, and Ke~Yi.
\newblock {Semantics of Ranking Queries for Probabilistic Data and Expected
  Ranks}.
\newblock In {\em 2009 IEEE 25th International Conference on Data Engineering},
  pages 305--316. IEEE, 2009.

\bibitem{li2010ranking}
Jian Li and Amol Deshpande.
\newblock {Ranking Continuous Probabilistic Datasets}.
\newblock {\em Proceedings of the VLDB Endowment}, 3(1-2):638--649, 2010.

\bibitem{hua2008ranking}
Ming Hua, Jian Pei, Wenjie Zhang, and Xuemin Lin.
\newblock Ranking queries on uncertain data: a probabilistic threshold
  approach.
\newblock In {\em Proceedings of the 2008 ACM SIGMOD international conference
  on Management of data}, pages 673--686, 2008.

\bibitem{yi2008efficient}
Ke~Yi, Feifei Li, George Kollios, and Divesh Srivastava.
\newblock Efficient processing of top-k queries in uncertain databases with
  x-relations.
\newblock {\em IEEE transactions on knowledge and data engineering},
  20(12):1669--1682, 2008.

\bibitem{bernecker2010scalable}
Thomas Bernecker, Hans-Peter Kriegel, Nikos Mamoulis, Matthias Renz, and
  Andreas Zuefle.
\newblock {Scalable Probabilistic Similarity Ranking in Uncertain Databases}.
\newblock {\em IEEE Transactions on Knowledge and Data Engineering},
  22(9):1234--1246, 2010.

\bibitem{LiSahDes11}
Jian Li, Barna Saha, and Amol Deshpande.
\newblock A unified approach to ranking in probabilistic databases.
\newblock {\em VLDB Journal}, 20(2):249--275, 2011.

\bibitem{bernecker2011novel}
Thomas Bernecker, Tobias Emrich, Hans-Peter Kriegel, Nikos Mamoulis, Matthias
  Renz, and Andreas Z{\"u}fle.
\newblock A novel probabilistic pruning approach to speed up similarity queries
  in uncertain databases.
\newblock In {\em 2011 IEEE 27th International Conference on Data Engineering},
  pages 339--350. IEEE, 2011.

\bibitem{hubig2012continuous}
Nina Hubig, Andreas Z{\"u}fle, Tobias Emrich, Mario~A Nascimento, Matthias
  Renz, and Hans-Peter Kriegel.
\newblock Continuous probabilistic sum queries in wireless sensor networks with
  ranges.
\newblock In {\em International Conference on Scientific and Statistical
  Database Management}, pages 96--105. Springer, 2012.

\end{thebibliography}
%
%



\end{document}